\documentclass[aps,pra,twocolumn,groupedaddress,showpacs,showkeys]{revtex4}

\usepackage{graphicx}
\begin{document}

\newcommand{\be}{\begin{equation}}
\newcommand{\ee}{\end{equation}}

\title{The Bohr-van Leeuwen theorem and the   thermal Casimir effect for conductors}

\author{ Giuseppe Bimonte}
\email[Bimonte@na.infn.it]
\affiliation{Dipartimento di Scienze Fisiche Universit\`{a} di
Napoli Federico II Complesso Universitario MSA, Via Cintia
I-80126 Napoli Italy and INFN Sezione di Napoli, ITALY\\
}

\date{\today}

\begin{abstract}
The problem of estimating the thermal corrections to   Casimir and
Casimir-Polder interactions in systems involving conducting plates
has attracted considerable attention in the recent literature on
dispersion forces. Alternative theoretical models, based on
distinct low-frequency extrapolations of the plates reflection
coefficient for transverse electric (TE) modes, provide widely
different predictions for the magnitude of this correction. In
this paper we examine the most widely used prescriptions for this
reflection coefficient from the point of view of their consistency
with the Bohr-van Leeuwen theorem of classical statistical
physics, stating that at thermal equilibrium transverse
electromagnetic fields decouple from matter in the classical
limit. We find that the theorem is satisfied if and only if the TE
reflection coefficient vanishes at zero frequency in the classical
limit.  This criterion appears to rule out some of the models that
have been considered recently for describing the thermal
correction to the Casimir pressure with   non-magnetic metallic
plates.

\end{abstract}


\maketitle

\section{INTRODUCTION}

Dispersion forces, frequently called "van der Waals" or
"molecular" forces, are long-range electromagnetic forces arising
from quantum and thermal charge and current fluctuations existing
in  microscopic or macroscopic bodies at thermal equilibrium. In
view of their pervasive role, from biology to chemistry, from
physics to engineering \cite{parse}, these weak forces have been
the subject of intense theoretical and experimental
investigations. A distinctive feature of dispersion forces is
intimately related to their long-range character. In fact, while
short-range  forces, like the exchange electromagnetic
interaction,  are determined by the particular microscopic
electronic structure of atoms and molecules, long-range forces
display a universal behavior at large distances. This feature is
clearly reflected in the famous theory of dispersion forces
developed long ago by Lishitz \cite{lifs}, where the particular
features of the bodies participating to the interaction, whether
atoms or macroscopic bodies, can be fully taken into account by
means of their macroscopic permittivities or polarizabilities.

After over fifty years, Lifshitz theory  still constitutes the
basic theoretical tool universally used by researchers in the
field, to interpret the results of modern experiments on
dispersion forces. As an example, we mention the numerous recent
experiments on the Casimir effect (for a recent review, see
\cite{Mohid}), and the beautiful new experiments on the
Casimir-Polder interaction of a Bose-Einstein condensate with a
dielectric substrate \cite{cornell}.  It is important to note that
the precision of the most recent experiments is of a such a level
that in order to correctly interpret the data it is now necessary
to take into full account a number of small corrections, like
temperature effects, the effect of surface corrugations, patch
effects   etc. (see Ref. \cite{Mohid} for details). In particular,
the necessity of careful electrostatic calibrations in precision
measurements of the Casimir force has been recently emphasized
\cite{iannuzzi}.

In this paper we focus our attention on the influence of material
properties of the bodies constituting the system. As remarked
above,  within Lifshitz theory these properties  are fully
described by the macroscopic permittivities of the bodies
participating in the interaction. As it is well known, the  latter
quantities are complex functions of the frequency $\omega$ of the
electromagnetic field, as an effect of dispersive and absorptive
properties displayed by all real materials. Now,  in Lifshitz
theory the free-energy associated with  van der Waals forces
between condensed bodies is expressed by a sum of terms depending
on the reflection coefficients of the plates, evaluated at
(imaginary) Matsubara frequencies $\xi_n=2 \pi n k_B T/\hbar$,
where $k_B$ is Boltzmann constant, $T$ is the temperature, and the
discrete index $n$ in the sum runs form zero to infinity. As a
result, evaluation of Lifshitz formula requires knowledge of the
reflection coefficients over a wide range of frequencies,
extending from zero up to a few times the characteristic frequency
$\omega_c=c/(2 a)$ of the system, where $a$ is the characteristic
separation between the bodies. For typical separations involved in
present experiments, ranging from a few tens of nanometers up to a
few microns, the characteristic frequency falls somewhere from the
IR part of the spectrum, to the near UV. In order to obtain an
accurate theoretical prediction for the free-energy, the common
practice today is to rely on detailed optical data of the material
used in the experiment. The importance of using accurate optical
data has been recently emphasized by the authors of Ref.
\cite{piro}, where it is shown that uncertainty in the optical
data may easily result in an uncertainty of several percent in the
evaluation of Lifshitz formula. One point should however be
stressed: for the purpose of evaluating the $n=0$ term of Lifshitz
formula, it is always necessary to make an extrapolation of the
reflection coefficients to zero frequency. This is a very delicate
point, indeed, because the result strongly depends on how the
extrapolation is done. Understanding what is the correct
extrapolation is crucial in particular for determining the thermal
correction to the free-energy in systems involving one or more
conductors, because this   correction is strongly affected by the
magnitude of the $n=0$ Matsubara term for transverse electric (TE)
polarization. This crucial term is determined by the TE reflection
coefficients of the plates at zero frequency, and as of now
different recipes have been proposed in the literature for this
coefficient, resulting in drastically different predictions for
the magnitude of the thermal correction. The most popular
prescriptions used currently to model real metals can be grouped
in two classes: Drude-like models and plasma-like models. The
former models \cite{sernelius} are characterized  by permittivity
functions $\epsilon(\omega)$ displaying at low frequency the same
$\omega^{-1}$ singularity of the familiar Drude model for the
permittivity of a ohmic conductor, resulting in a {\it vanishing}
TE reflection coefficient at zero frequency. On the contrary,
plasma-like  models are characterized by the  $\omega^{-2}$
singularity displayed  by the plasma model of IR optics,
extrapolated to zero frequency. The latter class of models
includes in particular the so-called generalized plasma model
considered in Ref. \cite{geyer}. In this model, ohmic dissipation
is not accounted for, and only dissipation associated with
interband transitions of core electron is considered. The
$\omega^{-2}$ singularity of plasma-like models entails a {\it
non-vanishing} TE reflection coefficient in the limit of zero
frequency, differently from Drude-like models. The different
behaviors of the TE reflection coefficient at zero frequency
implied by the two classes of models have a dramatic impact on the
Casimir force in the limit of large separations between the
plates, since Drude-like models lead to a force that is about
one-half that implied by plasma-like models. We remark also that
in this limit plasma-like models  give almost the same force as
the simple ideal-metal model of real conductors
\cite{brevik,qfext,Mohid}. At separations between the plates
smaller than one micron, where the Casimir force can be measured
most accurately, the various prescriptions imply predicted Casimir
forces that differ only by a few percents, and it is therefore
difficult to distinguish between them by an experiment.

The experimental situation is still unclear, because the present
accuracy of Casimir-force measurements is still not sufficient to
detect the temperature correction to the Casimir force between two
metallic bodies, and indeed there are at present several ongoing
and planned experiments to measure it \cite{brown}. A recent
accurate experiment using a micromechanical oscillator
\cite{decca} appears to favor the plasma model prescription, but
this claim is  not yet universally accepted by the community
\cite{brevik}. We remark that as of now there is only one
experiment that detected the influence of temperature on the
Casimir-Polder interaction of a Bose-Einstein condensate with a
dielectric substrate \cite{cornell}. Clarifying this problem is
important also in view of the many experiments on non-newtonian
forces at the submicron scale, some of which use metallic surfaces
at room temperature \cite{decca}, making it necessary to estimate
accurately the contribution of dispersion forces that must be
subtracted from the observed signal. For a lucid description of
the problems involved in precision Casimir experiments we address
the reader to the recent paper \cite{Iannuzzi}.

In this paper we shall examine the   low frequency prescriptions
for the TE reflection coefficient  of insulators and conductors,
generally used in investigations of dispersion forces, from the
point of view of their compatibility with a very well known
theorem of classical solid state physics, namely the Bohr-van
Leeuwen theorem \cite{van}. This theorem originated early in the
20th century, in an attempt to explain the absence of strong
diamagnetism in normal conductors placed in an external magnetic
field. After a simple physical argument due to Bohr, who showed
that no net diamagnetic currents can arise in a bounded conductor
subjected to a static magnetic field, the theorem was put on a
firm theoretical basis by H. J. van Leeuwen. In essence the
theorem states that in classical systems at thermal equilibrium
matter decouples from the transverse electromagnetic field. It
occurred to us that perhaps this theorem could be used to
discriminate between existing models used in the current
literature to describe dispersion forces. Let us see briefly how
the connection arises. One observes that the reflection
coefficients of a surface determine the macroscopic response of
the surface to an external electromagnetic probe placed outside
the surface. It is now the essence of the famous fluctuation
dissipation theorem \cite{callen} that such response functions are
intimately related to equilibrium averages of suitable macroscopic
observables of the system. In the case of interest to us, the
fluctuation dissipation theorem relates averages of the
fluctuating electromagnetic fields outside the surface to its
reflection coefficients \cite{agarwal}. This fundamental relation
is known to imply a set of general constraints, originating from
microscopic reversibility \cite{Casimir2}, that must be satisfied
by the reflection coefficients of any real material, like for
example important reciprocity relations \cite{santa}. Therefore we
were led to wonder if the Bohr-van Leeuwen theorem can be used to
put any further constraints on the permitted behavior of the
reflection coefficients. We remark that the Casimir and
Casimir-Polder interactions are equilibrium phenomena, and
therefore they must conform to the principles of equilibrium
statistical physics. In order to put this idea  to a test we shall
evaluate the spectrum of the fluctuating electromagnetic field in
the empty space outside  one slab, and between two plane-parallel
slabs, characterized by a local dielectric response, carefully
separating the longitudinal and the transverse components of the
electromagnetic field. Having done this, we shall verify whether
or not the transverse component of the field decouples from the
slab(s) in the classical limit, as required by the Bohr-van
Leeuwen theorem. Interestingly, we shall see that the answer
depends exclusively on the behavior of the reflection coefficients
of the slab(s), in the limit of zero frequency. In this way we
obtain a rather stringent test to decide whether a definite model
is admissible or not from the point of view of classical
statistical physics. The important result is that the Bohr-van
Leeuwen theorem is satisfied if and only if the TE reflection
coefficient vanishes at zero frequency, in the classical limit. As
a result, we find that Drude-like models are compatible with the
Bohr-van Leeuwen theorem, while neither plasma-like models nor the
ideal-metal model pass the test. Our conclusions appear to be
consistent with the findings of a recent paper \cite{Martin} which
presented a microscopic calculation of the Casimir force between
two metallic plates, in the asymptotic limit of large separations
between the plates, when the force is dominated by classical
thermal fluctuations and the theorem is supposed to apply. It was
found there that in this asymptotic regime, the microscopic model
assumed in \cite{Martin} predicts the same Casimir force as the
Drude prescription,  and therefore it is in disagreement with both
plasma-like models and the ideal metal results, which we recall
both predict in this limit a force of double magnitude as that
implied by Drude-like models.

We point out that in the recent literature on the thermal Casimir
effect, another criterion based on statistical physics has been
widely considered to discriminate between alternative models for
the reflection coefficients of a material slab. This other
criterion requires that the Nernst heat theorem be satisfied, in
the limit of zero temperature \cite{bezerra}.   This alternative
criterion leads to conclusions that are not in agreement with what
we found on the basis of the Bohr--van Leeuwen criterion, for one
finds that the Nernst heat theorem is not satisfied by the Drude
prescription in the idealized case of perfect crystals with no
defects, but it is satisfied both by the (generalized) plasma
models, and by the ideal metal model. In the Drude case, Nernst
theorem is however restored if an arbitrarily small amount of
impurities are present in the crystal \cite{sernelius2}.  While we
cannot offer a complete resolution of this contradiction, we
remark that since the Nernst theorem is intrinsically a quantum
result, this criterion is a sense orthogonal to the one proposed
in this paper, which is essentially classical. In our judgement,
in the absence of a definitive answer, we note that the Bohr--van
Leeuwen criterion appears to be more pertinent than the Nernst
criterion, for a theoretical assessment of the debated problem of
thermal corrections to the Casimir and Casimir-Polder effects at
room temperatures, since it is well known that the difficulties
posed by this problem are basically of a classical nature (see
Sec. V below). In addition, we note that the extrapolation to zero
temperature of Lifshitz theory poses very non-trivial problems,
associated for example with the possible presence of spatial
non-locality (anomalous skin effect \cite{svetovoy}). For recent
reviews of some theoretical aspects involved by the Nernst heat
theorem in the context of Casimir physics, we address the reader
to  Refs. \cite{brevik, elling}.

The paper is organized as follows. In Section II we review the
fluctuation-dissipation theorem, in the context of general linear
response theory, while in Section III the theorem is used to
derive expressions for the correlators of the electromagnetic
fields outside dielectrics and conductors. In Section IV we verify
if these correlators satisfy the Bohr- van Leeuwen theorem outside
a planar slab, for a number of models of dielectrics and
conductors. The case of a plane-parallel cavity, of the type used
in Casimir experiments, is considered in Section V. Finally,
Section VI contains our conclusions and a discussion of the
results.

\section{FLUCTUATION-DISSIPATION THEOREM}

In this Section we briefly review the principal results of linear
response theory, and in particular we present the general
fluctuation-dissipation theorem for linear dissipative media. For
a review of linear-response theory we address the reader to
Refs.\cite{callen}.

In linear-response theory, one considers a quantum-mechanical
system, characterized by a (time-independent) Hamiltonian $H_0$,
in a state of thermal equilibrium described by the density matrix
$\rho$ \be \rho=e^{-\beta H}/{\rm tr}(e^{-\beta H})\;,\ee where
$\beta=1/(k_B\,T)$. The system is then perturbed by an external
perturbation of the form: \be H_{\rm ext}=-\int d^3{\bf r} \sum_j
Q_j({\bf r},t)\,f_j({\bf r},t)\,\ee where $f_j({\bf r},t)$ are
external classical forces, and $Q_j({\bf r},t)$ is the dynamical
variable of the system conjugate to the force $f_j({\bf r},t)$.
One may assume without loss of generality that, in the absence of
external forces, the equilibrium values  of the quantities
$Q_j({\bf r},t)$ all vanish: $\langle Q_j({\bf r},t)\rangle=0$.
The presence of the external forces causes a deviation $\delta
\langle Q_i({\bf r},t)\rangle$ of the expectation values of
$Q_j({\bf r},t)$ from their equilibrium values. If the forces
$f_j({\bf r},t)$ are sufficiently weak, $\delta \langle Q_i({\bf
r},t)\rangle$ can be taken to be linear functionals of the applied
forces $f_j({\bf r},t)$ according to the formula: \be \delta
\langle Q_i({\bf r},t)\rangle=\sum_j \int d^3{\bf r}
\int_{-\infty}^t dt' \phi_{ij}({\bf r},{\bf r}',t-t')\,f_j({\bf
r}',t')\;.\ee The above Equation assumes that the system was in
equilibrium at $t=-\infty$, and that it reacts to the  external
force in a causal way. The quantities $\phi_{ij}({\bf r},{\bf
r}',t-t')$ are called response functions of the system. In
principle, they can be measured by applying to the system of
interest suitable external classical probes. We now define the
admittance ${\tilde \phi}_{ij}({\bf r},{\bf r}',\omega)$ as the
(one-sided) Fourier transform of the response function
${\phi}_{ij}({\bf r},{\bf r}',t)$: \be {\tilde \phi}_{ij}({\bf
r},{\bf r}',\omega)=\int_0^{\infty} dt \, \phi_{ij}({\bf r},{\bf
r}',t) \, e^{i \omega t}\;.\label{adm}\ee Being the one-sided
transform of a real quantity, the admittance satisfies distinctive
analyticity and reality properties. First, it is an analytic
function of the complex frequency $w$ in the upper complex plane
${\cal C}^+\equiv \{w=\omega+i \delta,\;\delta>0\}$. Second, it
satisfies in ${\cal C}^+$ the following reality condition \be
{\tilde \phi}_{ij}({\bf r},{\bf r}',-w^*)={\tilde
\phi}_{ij}^*({\bf r},{\bf r}',w)\;.\label{real} \ee The latter
property in particular implies that the admittance is real along
the imaginary frequency axis ${\tilde \phi}_{ij}({\bf r},{\bf
r}',i \xi)={\tilde \phi}_{ij}^*({\bf r},{\bf r}',i\xi)$.

By a straightforward computation in time-dependent perturbation
theory one may prove that the response functions $\phi_{ij}({\bf
r},{\bf r}',t-t')$  are related to the equilibrium  expectation
values  of the commutators of the dynamical variables $Q_i({\bf
r},t)$,  in the absence of external forces: \be \phi_{ij}({\bf
r},{\bf r}',t-t')=\Delta_{ij}({\bf r},{\bf
r}',t-t')\;\theta(t-t').\label{resp}\ee Here $\theta(x)$ is
Heaviside step function ($\theta(x)=1$ for $x > 0$, $\theta(x)=0$
for $x<0$) and \be \Delta_{ij}({\bf r},{\bf r}',t-t')=\frac{i}{
\hbar}\,\langle [Q_i({\bf r},t),Q_j({\bf
r}',t')]\rangle\;,\label{resp2}\ee with $Q_i({\bf r},t)$   the
Heisenberg operator: \be Q_i({\bf r},t)=e^{i H_0 t/\hbar} Q_i({\bf
r},0)e^{-i H_0 t/\hbar}\;.\ee
As it is well known, starting from Eq. (\ref{resp}) it is possible
to derive several general fluctuation-dissipation theorems, that
allow to express the (symmetrized) correlation functions of the
quantities $Q_i({\bf r},t)$ in terms of the dissipative component
of the response functions $\phi_{ij}$ \cite{callen}. The form of
the fluctuation-dissipation theorem of interest to us is expressed
by the following relation $$ \int_{-\infty}^{\infty} dt
\,\Delta_{ij}({\bf r},{\bf r}',t)\,e^{i \omega t}$$
\be=\frac{i\,\omega}{E_{\beta}(\omega)} \int_{-\infty}^{\infty}
dt\, \langle \{ Q_i({\bf r},t)\,Q_j({\bf r}',0)\}\rangle e^{i
\omega t}\,\label{flucth}\ee where
$$
E_{\beta}(\omega)=\frac{\hbar \omega}{2} \coth \left(\frac{\hbar
\omega}{2 k_B T}\right)
$$
is the average free-energy of a quantum oscillator with frequency
$\omega$ in equilibrium at temperature $T$, and $\{AB\}=(AB+BA)/2$
denotes  the symmetrized  product of the operators $A$ and $B$. If
$\Delta_{ij}({\bf r},{\bf r}',t)$ has a definite parity under
inversion of time, we can easily verify that the l.h.s. of Eq.
(\ref{flucth}) can be expressed in terms of the admittance.
Consider first the case when $\Delta_{ij}({\bf r},{\bf r}',t)$ is
odd in $t$. This is the case of interest to us, because the
commutators of the electromagnetic fields are indeed odd in time.
Then we find:
$$ \int_{-\infty}^{\infty} dt \,\Delta_{ij}({\bf r},{\bf
r}',t)\,e^{i \omega t}= 2 \,i \int_{0}^{\infty}
dt\,\Delta_{ij}({\bf r},{\bf r}',t)\,\sin(\omega t)$$ \be
=2\,i\,{\rm Im}[{\tilde \phi}_{ij}({\bf r},{\bf
r}',\omega)]\;.\label{odd}\ee  Upon substituting the expression on
the second line  of Eq. (\ref{odd})  in the l.h.s. of Eq.
(\ref{flucth}), and after performing the inverse Fourier transform
of both sides, we then obtain the following important relation:
$$
\langle\{ Q_i({\bf r},t)\,Q_j({\bf r}',0)\}\rangle
$$
\be = \frac{1}{\pi} \int_{-\infty}^{\infty}\frac{d \omega}{\omega
}\, E_{\beta}(\omega)\, {\rm Im}[{\tilde \phi}_{ij}({\bf r},{\bf
r}',\omega)]\, e^{-i \omega t}\;.\ee  By further exploiting the
fact that, by virtue of Eq. (\ref{real}), the imaginary part of
the admittance is an odd function of $\omega$, we can rewrite the
above Equation as:
$$
\langle\{ Q_i({\bf r},t)\,Q_j({\bf r}',0)\}\rangle
$$
\be =\frac{2}{\pi} \int_{0}^{\infty}\frac{d \omega}{ \omega}\,
E_{\beta}(\omega)\, {\rm Im}\,[{\tilde \phi}_{ij}({\bf r},{\bf
r}',\omega)] \cos(\omega t)\;.\label{flucodd}\ee Even if we shall
not need it, for completeness we report also the analogous
relation that holds if $\Delta_{ij}({\bf r},{\bf r}',t)$ is even
in $t$:
$$
\langle\{ Q_i({\bf r},t)\,Q_j({\bf r}',0)\}\rangle
$$
\be =\frac{2}{\pi}\, \int_{0}^{\infty}\frac{d \omega}{
\omega}\,E_{\beta}(\omega)\, {\rm Re}\,[{\tilde \phi}_{ij}({\bf
r},{\bf r}',\omega)] \sin(\omega t)\;.\label{fluceven}\ee The
above two relations constitute the content of the
fluctuation-dissipation theorem. Note in particular that in the
odd case the two-times correlation functions of the quantities $
Q_i({\bf r},t)$ in Eq. (\ref{flucodd}) is expressed in terms of
the dissipative part of the admittance. It is important to remark
that the integrands on the r.h.s. of Eqs. (\ref{flucodd}) and
(\ref{fluceven}) have no singularity at $\omega=0$, despite the
presence of the singular factor $\omega^{-1}$. This so because,
from the definition Eq. (\ref{adm}), we see that for vanishing
frequency the admittance ${\tilde \phi}_{ij}({\bf r},{\bf r}',0)$
is a finite and real quantity. This ensures that the integrands in
Eqs. (\ref{flucodd}) and (\ref{fluceven}) both have finite limits
as $\omega$ approaches zero.

In view of a next use, it is now useful to derive a formula for
the equal-time correlators of the quantities $ Q_i({\bf r},t)$, in
the classical limit. This can be easily done by setting $t=0$ and
taking the limit $\hbar \rightarrow 0$ in the r.h.s. of Eqs.
(\ref{flucodd}) and (\ref{fluceven}). The only non-trivial case to
consider is the odd one, for in the even case we see from Eq.
(\ref{fluceven}) that the equal-time correlators vanish
identically. Then, from Eq. (\ref{flucodd}) we obtain:
$$
\lim_{\hbar \rightarrow 0}\langle\{ Q_i({\bf r},0)\,Q_j({\bf
r}',0)\}\rangle
$$
\be =\frac{2 k_B T}{\pi} \,{\rm Im}\,\int_{0}^{\infty}\frac{d
\omega}{ \omega}\,  {\tilde \phi}_{ij}({\bf r},{\bf r}',\omega)
\;.\label{rot}\ee Assuming that the admittance ${\tilde
\phi}_{ij}({\bf r},{\bf r}',w)$ vanishes sufficiently fast at
complex infinity in ${\cal C}^+$, we can take advantage of
analyticity in ${\cal C}^+$ of    ${\tilde \phi}_{ij}({\bf r},{\bf
r}',w)$ to evaluate the integral on the r.h.s., by rotating the
integration contour from the real axis to the imaginary one. The
rotated contour of integration $\Gamma$ consists of an
infinitesimal arc surrounding the origin in the right sector of
${\cal C}^+$, followed by the whole imaginary axis. Since the
admittance is real along the imaginary axis, it is clear that the
part of the integral over $\Gamma$ extending along the imaginary
axis does not contribute to the r.h.s. of Eq.(\ref{rot}), and
therefore we find that the   imaginary part of the integral
results entirely from the contribution of the infinitesimal arc
surrounding the origin. After easy evaluation of the latter
contribution, we obtain the simple result: \be \lim_{\hbar
\rightarrow 0}\langle\{ Q_i({\bf r},0)\,Q_j({\bf r}',0)\}\rangle
=k_B T \,{\tilde \phi}_{ij}({\bf r},{\bf r}',0)\;.\label{clas}\ee
We then reach the important conclusion that in the classical limit
the equal-time correlators for the quantities $Q_i({\bf r})$ are
simply proportional to the zero-frequency limit of the admittance.
It is opportune to remark that in the above derivation  we have
implicitly assumed that the admittance itself is independent of
$\hbar$.  Obviously when quantum effects contribute to the
admittance, the quantity ${\tilde \phi}_{ij}({\bf r},{\bf r}',0)$
in the r.h.s. of Eq. (\ref{clas}) must be understood as the
classical limit of the admittance at zero-frequency.

\section{CORRELATORS OF THE ELECTROMAGNETIC FIELD OUTSIDE DIELECTRICS AND CONDUCTORS}

In this Section we shall use the methods described in the previous
Section to derive formulae for the correlators of the
electromagnetic fields outside dielectrics and conductors. In the
spirit of linear response theory, this is done by placing outside
the bodies a suitable distribution of classical electric and
magnetic dipoles, that work as external probes for the
electromagnetic field \cite{agarwal}. In order to separate the
longitudinal and the transverse parts of the field, we consider an
external Hamiltonian of the following form \cite{bimonte} \be
H_{\rm ext}=\int d^3{\bf r}[U({\bf r},t) \rho^{({\rm ext})} ({\bf
r},t)-\frac{1}{c} {\bf A}_{\perp}({\bf r},t)\cdot {\bf j}^{({\rm
ext})}_{\perp} ({\bf r},t) ]\;, \label{Hextfin}\ee where $U({\bf
r},t)$ and ${\bf A}_{\perp}({\bf r},t)$ are, respectively, the
scalar and the transverse vector potentials for the
electromagnetic field, while $\rho^{({\rm ext})} ({\bf r},t)$ and
${\bf j}^{({\rm ext})}_{\perp} ({\bf r},t)$ denote, respectively,
external classical distributions of charge and current. Note  that
${\bf j}^{({\rm ext})}_{\perp} ({\bf r},t)$ is assumed to be
transverse: \be {\bf \nabla} \cdot {\bf j}^{({\rm ext})}_{\perp}
=0\;.\ee In what follows we shall assume that the bodies
constituting the system are non-magnetic ($\mu=1$) dielectrics or
conductors with sharp boundaries, characterized each by a
frequency dependent electric permittivity $\epsilon(\omega)$. It
is moreover assumed that the bodies are homogeneous, in such a way
that the permittivities are constant functions of the position
within the volume occupied by each body. Under such conditions, it
is shown in Ref.\cite{bimonte} that we have two independent sets
of response functions for the scalar and the vector potential: \be
{U}({\bf r},t)=\int_{-\infty}^{t} d t' \int d^3 {\bf r}'
{{G}}({\bf r},{\bf r}',t-t')\,{\rho}^{({\rm ext})}({\bf
r}',t')\;,\ee \be {{\bf A}}_{\perp}({\bf
r},t)=\frac{1}{c}\int_{-\infty}^{t}  d t'\int d^3 {\bf r}' { {\bf
G}}_{\perp}({\bf r},{\bf r}',t-t')\cdot{ {\bf j}}_{\perp}^{({\rm
ext})}({\bf r}',t')\;,\ee where ${ {\bf G}}_{\perp}({\bf r},{\bf
r}',t-t')$ has to be understood as a dyadic Green function.
Recalling that the commutators of the electromagnetic fields are
odd functions of time, from the general fluctuation-dissipation
theorem Eq. (\ref{flucodd}) we   obtain the following expressions
for the correlators of the electromagnetic fields:
$$
\langle\{U({\bf r},t)\,U({\bf r}',0)\}\rangle
$$
\be =-\frac{2}{\pi} \int_{0}^{\infty}\frac{d
\omega}{\omega}\,E_{\beta}(\omega)\,{\rm Im}\,[{\tilde G}({\bf
r},{\bf r}',\omega)] \cos(\omega t)\;,\label{UUzero}\ee \be
\langle\{U({\bf r},t)\,A_{\perp i}({\bf r}',0)\}\rangle =0\;,\ee
$$
\langle\{A_{\perp i}({\bf r},t)\,A_{\perp j}({\bf r}',0)\}\rangle
$$
\be = \frac{2}{\pi} \int_{0}^{\infty}\frac{d
\omega}{\omega}\,E_{\beta}(\omega)\, {\rm Im}\,[{\tilde G}_{\perp
ij}({\bf r},{\bf r}',\omega)] \cos(\omega t)\;,\label{AAzero}\ee
where ${\tilde G}({\bf r},{\bf r}',\omega)$ and ${\tilde {\bf
G}}_{\perp}({\bf r},{\bf r}',\omega)$ denote the one-sided Fourier
transforms of the Greens functions: \be {\tilde G}({\bf r},{\bf
r}',\omega)=\int_0^{\infty} dt \, {G}({\bf r},{\bf r}',t)\,e^{i
\omega t}\;,\ee \be {\tilde {\bf G}}_{\perp}({\bf r},{\bf
r}',\omega)=\int_0^{\infty} dt \, {\bf G}_{\perp}({\bf r},{\bf
r}',t)\,e^{i \omega t}\;.\ee The Fourier transforms of the Green's
functions can be obtained by solving the following field Equations
implied by  macroscopic Maxwell Equations: \be {\bf \nabla}\cdot
[\,\epsilon({\bf r},\omega) \,{\bf \nabla} \tilde{G}\,]=-4 \pi
\,\delta({\bf r}-{\bf r'})\;,\label{diffsca}\ee \be (\triangle +
\,\epsilon({\bf r},\omega)\,\omega^2/c^2){\tilde {\bf
G}}_{\perp}({\bf r},{\bf r}',\omega)=- {4 \pi} \,{\bf
\delta}_{\perp}({\bf r}-{\bf r'})\;,\label{difften}\ee where ${\bf
\delta}_{\perp}({\bf r}-{\bf r'})$ is the transverse
delta-function dyad: \be \delta_{ij}^{\perp}({\bf x})=\int{d^3
{\bf k}} \left(\delta_{ij}-\frac{k_i k_j}{k^2} \right)\,e^{i{\bf
k} \cdot {\bf x}}\,,\label{delta}\ee with $k=|{\bf k}|$. The above
field equations must be supplemented by standard boundary
conditions
  at the bodies interfaces, and must be further subjected to
the conditions required for retarded Green's functions. The
Green's functions ${\tilde G}$ and ${\tilde {\bf G}}_{\perp}$
satisfy  a number of general properties, that are consequences of
microscopic reversibility and of analyticity and reality
properties of the permittivity $\epsilon(\omega)$ of any causal
material. For a review of these important properties we address
the reader to Ref. \cite{bimonte} (and Refs. therein).

For our purposes, it is convenient to split the Green's functions,
{\it outside} the bodies, as sums of an empty-space contribution
plus a correction arising from the material bodies: \be {  { G}}
({\bf r},{\bf r}',t-t')= {{ G}}^{(0)}({\bf r}-{\bf
r}',t-t')+\,{{F}}^{(\rm mat)}({\bf r},{\bf
r}',t-t')\;,\label{decscagen}\ee and \be {{\bf G}}_{\perp}({\bf
r},{\bf r}',t-t')={{\bf G}}^{(0)}_{\perp}({\bf r}-{\bf r}',t-t')+{
{\bf F}}^{(\rm mat)}_{\perp}({\bf r},{\bf
r}',t-t')\;.\label{dectengen}\ee Here, ${{ G}}^{(0)}$ and ${{\bf
G}}^{(0)}_{\perp}$ denote the Green's functions in free space,
while ${{F}}^{(\rm mat)}$ and ${{\bf F}}^{(\rm mat)}_{\perp}$
describe the effects resulting from the presence of the bodies.
Such a splitting presents the advantage that all singularities are
included in the free parts ${{ G}}^{(0)}$ and ${{\bf
G}}^{(0)}_{\perp}$, while the quantities ${{F}}^{(\rm mat)}$ and
${{\bf F}}^{(\rm mat)}_{\perp}$ are smooth ordinary functions of
${\bf r}$ and ${\bf r}'$. Upon using Eqs. (\ref{decscagen}) and
(\ref{dectengen}) into the r.h.s. of Eqs.
(\ref{UUzero}-\ref{AAzero}) we obtain the following equations for
the {\it changes} of the field correlators outside the material
bodies, arising from the presence of the bodies:
$$
\delta\langle\{U({\bf r},t)\,U({\bf r}',0)\}\rangle
  $$
\be =- \frac{2}{\pi} \int_{0}^{\infty}\frac{d
\omega}{\omega}\,E_{\beta}(\omega)\, {\rm Im}\,[{\tilde F}^{(\rm
mat)}({\bf r},{\bf r}',\omega)] \cos(\omega t)\;,\label{UU}\ee \be
\delta\langle\{U({\bf r},t)\,A_{\perp i}({\bf r}',0)\}\rangle
=0\;,\label{UA}\ee
$$
\delta\langle\{A_{\perp i}({\bf r},t)\,A_{\perp j}({\bf
r}',0)\}\rangle
$$
\be = \frac{2}{\pi} \int_{0}^{\infty}\frac{d
\omega}{\omega}\,E_{\beta}(\omega)\, {\rm Im}\,[{\tilde F}_{\perp
ij}^{(\rm mat)}({\bf r},{\bf r}',\omega)] \cos(\omega
t)\;,\label{AA}\ee with an obvious meaning for the symbols. It is
now easy to derive from the above formulae the expressions for the
corresponding changes of the equal-time correlators of the
electromagnetic fields that we shall need in the following
Sections. For the longitudinal electric field ${\bf E}_{\|}=-{\bf
\nabla} U$, from Eq. (\ref{UU}) we obtain
$$
\delta\langle\{E_{\| i}({\bf r},0)\,E_{\| j}({\bf r}',0)\}\rangle
$$
\be =-\frac{2}{\pi} \int_{0}^{\infty}\frac{d
\omega}{\omega}\,E_{\beta}(\omega)\, {\rm Im}\,\left(\,
\frac{\partial^2 {\tilde F}^{(\rm mat)}}{\partial x_i
\partial x'_j}\right) \;,\label{ElEl}\ee
For the transverse electric field ${\bf E}_{\perp}$ and for the
magnetic field ${\bf B}$, since ${\bf E}_{\perp}=-c^{-1}\;
\partial {\bf A}_{\perp}/\partial t$, and ${\bf B}={\bf \nabla
\times A}_{\perp}$, from Eq. (\ref{AA}) we obtain:
$$
\delta\langle\{E_{\perp i}({\bf r},0)\,E_{\perp j}({\bf
r}',0)\}\rangle
$$
\be =\frac{2}{\pi} \int_{0}^{\infty}\frac{d
\omega}{\omega}\,E_{\beta}(\omega)\,k_0^2\, {\rm Im}\,[\,{\tilde
F}_{\perp ij}^{(\rm mat)}({\bf r},{\bf r}',\omega)]
\;,\label{EE}\ee where $k_0=\omega/c$ and
$$
\delta\langle\{B_{i}({\bf r},0)\,B_{j}({\bf r}',0)\}\rangle
$$
\be = \frac{2}{\pi} \int_{0}^{\infty}\frac{d
\omega}{\omega}\,E_{\beta}(\omega)\, {\rm Im}\,[\,({\bf {\stackrel
\rightarrow \nabla}_{\bf r} \times {\tilde F} _{\perp}^{(\rm
mat)}\times{\stackrel \leftarrow \nabla}_{\bf r'}})_{ij}]
\;.\label{BB}\ee  Finally, from Eq. (\ref{clas}) we obtain the
following equations for the matter contributions to the equal-time
electric and magnetic correlators in the classical limit: \be
\lim_{\hbar \rightarrow 0}\,\delta \langle\{E_{\perp i}({\bf
r},0)\,E_{\perp j}({\bf r}',0)\}\rangle = k_B \,T \,{\cal E}^{(\rm
cl)}_{\perp ij}({\bf r},{\bf r}')
  \;,\label{EEclas}\ee and
\be \lim_{\hbar \rightarrow 0}\,\delta \langle\{B_{\perp i}({\bf
r},0)\,B_{\perp j}({\bf r}',0)\}\rangle  = k_B \,T \,{\cal
B}^{(\rm cl)}_{\perp ij}({\bf r},{\bf r}')\;,\label{BBclas}\ee
where we defined \be {\cal E}^{(\rm cl)}_{\perp ij}({\bf r},{\bf
r}')= \lim_{\omega \rightarrow 0} (k_0^2\, \,{\tilde F}_{\perp
ij}^{(\rm mat)}({\bf r},{\bf r}',\omega))\,,\label{cale}\ee and
\be {\cal B}^{(\rm cl)}_{\perp ij}({\bf r},{\bf r}')=\lim_{\omega
\rightarrow 0}  \,({\bf {\stackrel \rightarrow \nabla}_{\bf r}
\times {\tilde F} _{\perp}^{(\rm mat)}\times{\stackrel \leftarrow
\nabla}_{\bf r'}})_{ij}\;. \label{calb}\ee According to the
Bohr-van Leeuwen theorem \cite{van} the transverse electromagnetic
fields decouples from matter in the classical limit. From Eqs.
(\ref{EEclas}) and (\ref{BBclas}) we see that the theorem is
fulfilled if and only if the quantities ${\cal E}^{(\rm
cl)}_{\perp ij}$ and ${\cal B}^{(\rm cl)}_{\perp ij}$ all vanish
identically outside the bodies. Equipped with these formulae, we
are ready now to examine whether the theorem is satisfied in the
simple case of a dielectric slab.

\section{THE BOHR-VAN LEEUWEN THEOREM FOR ONE DIELECTRIC OR CONDUCTING SLAB}

In this Section we use the results of the previous Sections to
verify whether the Bohr-van Leeuwen theorem  is satisfied outside
a plane-parallel dielectric or conducting slab characterized by a
spatially local permittivity $\epsilon(\omega)$. We choose our
cartesian coordinate system in such a way that the $z$ axis is
perpendicular to the slab surface, with the slab occupying the
$z<0$ half-space.

The relevant Green's functions for this problem have already been
worked out in Ref. \cite{bimonte}. The matter contribution
${\tilde {\bf F}}^{(\rm wall)}_{\perp}$ to the tensor Green's
functions  was found to have the following form: \be {\tilde {
F}_{\perp ij}}^{({\rm wall})}={\tilde {  U}}^{({\rm
wall})}_{ij}+{\tilde {  V}}^{({\rm wall})}_{ij}
\;.\label{Gwallsplit}\ee In this equation, ${\tilde {  U}}^{({\rm
wall})}_{ij}$ is the quantity
$$ {\tilde { U}_{ij}}^{({\rm wall})}=i \int \frac{d^2 {\bf
k}_{\perp}}{2 \pi\,{k_z}} \left( e_{\perp i} e_{\perp
j}\,r^{(s)}(\omega,{ k}_{\perp})\right.$$ \be
 +\left.\frac{\xi^{(+)}_i \xi^{(-)}_j}{k_0^2}\,r^{(p)}(\omega,{
k}_{\perp}) \right) \,e^{i {\bf k}^{(+)}\cdot {\bf r}-i{\bf
k}^{(-)}\cdot {\bf r}' }\;,\label{Ufin}\ee where ${\bf k}_{\perp}$
denotes the projection of the wave-vector onto the $(x,y)$ plane,
and we have defined $k_z=\sqrt{k_0^2-k_{\perp}^2}$ (the square
root is defined such that ${\rm Im}(k_z) > 0$), ${\bf
e_{\perp}}={\hat {\bf z}}\times {\hat {\bf k}}_{\perp}$, ${\bf
k}^{(\pm)}={\bf k}_{\perp}\pm k_z {\hat {\bf z}}$ and ${\bf
\xi}^{\pm}=k_{\perp}{\hat {\bf z}} \mp k_z {\hat {\bf
k}}_{\perp}$, while  $r^{(s)}(\omega,{\bf k}_{\perp})$ and
$r^{(p)}(\omega,{\bf k}_{\perp})$ are the familiar Fresnel
reflections coefficients for TE and TM waves, respectively: \be
r^{(s)}(\omega,{ k}_{\perp})=\frac{k_z-s}{k_z+s}\;,\label{rs}\ee
\be r^{(p)}(\omega,{ k}_{\perp})=\frac{\epsilon(\omega)\,
k_z-s}{\epsilon(\omega)\, k_z+s}\;,\label{rp}\ee where
$s=\sqrt{\epsilon(\omega) k_0^2-k_{\perp}^2}$, and again the
square root is defined such that ${\rm Im}(s)>0$. For the quantity
${\tilde { V}}^{({\rm wall})}_{ij}$ we have \be {\tilde {
V}}^{({\rm wall})}_{ij}=\frac{1}{k_0^2} \frac{\partial^2 {\tilde
\Psi}^{({\rm wall})}}{\partial x_i
\partial x'_j}\ee
where \be {\tilde {\Psi}}^{(\rm wall)}=- {\bar r}(\omega)\,\int
\frac{d^2 {\bf k}_{\perp}}{2 \pi\,k_{\perp}}\,  e^{i {\bar {\bf
k}} ^{(+)}\cdot {\bf r} -i {\bar {\bf k}}^{(-)} {\bf r} '
}\;,\;\;z \ge 0\,,\label{Fwall0}\ee where   ${\bar {\bf
k}}^{(\pm)}={\bf k}_{\perp} \pm i k_{\perp}\,{\hat {\bf z}}$ and
the reflection coefficient ${\bar r}(\omega)$ is \be {\bar
r}(\omega)=\frac{\epsilon(\omega)-1}{\epsilon(\omega)+1}\,.\label{refsca}\ee
We can now evaluate the quantities ${\cal E}^{(\rm cl)}_{\perp
ij}$ and ${\cal B}^{(\rm cl)}_{\perp ij}$ defined in Eqs.
(\ref{cale}) and (\ref{calb}). Using the following relations   \be
{\bf \xi}^{(\pm)}=\mp i\,{\bar {\bf k}}^{(\pm)}+ O(\omega^2)\;,\ee
\be {\bf k}^{(\pm)}= {\bar {\bf k}}^{(\pm)}+ O(\omega^2)\;,\ee \be
k_z=i\,k_{\perp}+ O(\omega^2)\;,\ee and observing that ${\bf
k}^{(\pm)} \times {\bf e_{\perp}}={\bf \xi}^{(\pm)}$ and ${\bf
\xi}^{(\pm)} \times {\bf k}^{(\pm)} =k_0^2\,{\bf e}_{\perp}$ it is
easy to verify that
$$ {\cal E}^{(\rm cl)}_{\perp ij}=\lim_{\omega \rightarrow 0}\int
\frac{d^2 {\bf k}_{\perp}}{2 \pi\,k_{\perp}} \left\{ k_0^2
\,r^{(s)}(\omega,{ k}_{\perp})\, e_{\perp i} e_{\perp
j}\,\right.$$\be \left.+ \,[r^{(p)}(\omega,{ k}_{\perp})-{\bar
r}(\omega)]\, {\bar { k}}_i^{(+)}{\bar {
k}}_j^{(-)}\frac{}{}\right\} e^{i {\bar {\bf k}} ^{(+)}\cdot {\bf
r} -i {\bar {\bf k}}^{(-)} {\bf r} ' }\;,\label{caleslab}\ee
$$ {\cal B}^{(\rm cl)}_{\perp ij}=\lim_{\omega
\rightarrow 0}\int \frac{d^2 {\bf k}_{\perp}}{2 \pi\,k_{\perp}}
\left\{ \,r^{(s)}(\omega,{ k}_{\perp})\,{\bar { k}}_i^{(+)}{\bar {
k}}_j^{(-)} \,\right.$$\be \left.+ \, k_0^2\,r^{(p)}(\omega,{
k}_{\perp})\,e_{\perp i} e_{\perp j}\, \frac{}{}\right\} e^{i
{\bar {\bf k}} ^{(+)}\cdot {\bf r} -i {\bar {\bf k}}^{(-)} {\bf r}
' }\;.\label{calbslab}\ee As we see from the above two Equations,
whether the Bohr-van Leeuwen theorem is satisfied or not depends
entirely on the behavior of the reflection coefficients, or what
is the same of the electric permittivity $\epsilon(\omega)$ of the
slab, in the limit of zero frequency. We consider now several
classes of models for the low-frequency behavior of the
permittivity of the materials that are usually used in present
experiments on dispersion forces \cite{Mohid}. First we have
insulators, whose permittivities approach a finite limit at zero
frequency: \be
\epsilon(\omega)=\epsilon_0+O(\omega)\;\;\;\;\;({\rm
insulator})\;.\label{ins}\ee Then we have  normal (i.e. non
superconducting) non-magnetic ohmic conductors. For these
materials several distinct models have been considered in the
current literature on dispersion forces, and we classify them
generally as Drude-like models, plasma-like models and the ideal
metal model. By Drude-like models we mean any models  displaying
at low frequency the same singular behavior of the familiar
Drude-model, characterized by an $\omega^{-1}$ singularity: \be
\epsilon(\omega)=\frac{4 \pi i
\sigma_0}{\omega}+O(1)\\;\;\;\;\;({\rm
Drude-like\;models})\;,\label{drude}\ee where $\sigma_0$ is the dc
conductivity. By plasma-like models we mean instead any models
characterized by the same  $\omega^{-2}$ singularity displayed by
the familiar plasma-model of IR optics \be
\epsilon(\omega)=-\frac{\Omega_P^2}{\omega^2}+O(\omega^{-1})\;\;\;\;\;({\rm
plasma-like\;models})\;,\label{plasma}\ee where $\Omega_P$ is the
plasma frequency. These models include in particular the so-called
generalized plasma model considered recently in connection with
the Casimir effect \cite{geyer}. Finally we have the ideal-metal
model which is better formulated directly in terms of the
reflection coefficients: \be r^{(p)}={\bar r}=-
r^{(s)}=1\;\;\;\;\;({\rm ideal\;metal})\;.\label{ideal}\ee We  now
estimate  the quantities ${\cal E}^{(\rm cl)}_{\perp ij}$ and
${\cal B}^{(\rm cl)}_{\perp ij}$ using the above models. It is a
simple matter to verify that for all models, the reflection
coefficients $r^{(p)}$, $r^{(s)}$ and ${\bar r}$ are finite in the
limit $\omega \rightarrow 0$. It is also possible to check (see
also Appendix B of Ref. \cite{bimonte}) that the difference
$r^{(p)}(\omega,{ k}_{\perp})-{\bar r}(\omega)$ occurring in the
r.h.s. of Eq. (\ref{caleslab}), vanishes always at zero frequency
(in the ideal case it is identically zero). In view of this, we
see from Eq. (\ref{caleslab}) that the quantities ${\cal E}^{(\rm
cl)}_{\perp ij}$ vanish for all models: \be {\cal E}^{(\rm
cl)}_{\perp ij}=0\;.\ee However, in the case of ${\cal B}^{(\rm
cl)}_{\perp ij}$ we see from Eq. (\ref{calbslab}) that only the
second term between the curly brackets vanishes always in the
limit of zero frequency, leaving us with: \be {\cal B}^{(\rm
cl)}_{\perp ij}= \int \frac{d^2 {\bf k}_{\perp}}{2
\pi\,k_{\perp}}\, r^{(s)}(0,{ k}_{\perp})\,{\bar {
k}}_i^{(+)}{\bar { k}}_j^{(-)}e^{i {\bar {\bf k}} ^{(+)}\cdot {\bf
r} -i {\bar {\bf k}}^{(-)} {\bf r} ' }\;,\label{calbslabbis}\ee
where we set \be r^{(s)}(0,{ k}_{\perp}) \equiv \lim_{\omega
\rightarrow 0} r^{(s)}(\omega,{
k}_{\perp})\;.\label{rTEzerofre}\ee This equation shows that the
key quantity to consider is the zero-frequency limit of the TE
reflection coefficient, for we see that the Bohr-van Leeuwen is
satisfied if and only if $ r^{(s)}(0,{ k}_{\perp})$ vanishes. Now,
it is easily seen that the insulator model and Drude-like models
both imply a vanishing value for $ r^{(s)}(0,{ k}_{\perp})$: \be
r^{(s)}(0,{ k}_{\perp})=0\;\;\;\;({\rm insul.\;and
\;Drude-like\;models}),\label{rTEins}\ee On the contrary, with
plasma-like models we find \be r^{(s)}(0,{
k}_{\perp})=\frac{k_{\perp}-\sqrt{k_{\perp}^2+\Omega_P^2/c^2}}{k_{\perp}+
\sqrt{k_{\perp}^2+\Omega_P^2/c^2}}\;\;\;\;({\rm
plasma-like\;models})\;.\label{rTEpl}\ee  For the ideal-metal
model we obviously have \be r^{(s)}(0,{
k}_{\perp})=-1\;\;\;\;({\rm ideal\;metal})\;.\label{rTEid}\ee

 In view of these formulae, we
reach the important conclusion that both the dielectric and the
ohmic (or Drude) models  are consistent with the Bohr-van Leeuwen
theorem, while both  plasma-like models and the ideal model are
not. It is interesting to remark that even when the latter two
models are considered, the quantity ${\cal B}^{(\rm cl)}_{\perp
ij}$ becomes negligible at large distances from the slab, and
therefore the inconsistency revealed here is expected to be
important only in the study of proximity effects like the Casimir
effect to be considered in the next Section.

To avoid misunderstandings, we should repeat the warning made at
the end Section II. We are tacitly admitting here that the
reflection coefficient $r^{(s)}(0,{ k}_{\perp})$ is independent of
$\hbar$. Obviously, when quantum effects are important, it is
understood that  in Eq. (\ref{calbslabbis}) one should take the
classical limit of $r^{(s)}(0,{ k}_{\perp})$. This consideration
applies for example both to magnetic materials, and especially so
to superconductors \cite{bimonte2}.

\section{THE BOHR-VAN LEEUWEN THEOREM AND THE THERMAL CASIMIR EFFECT}

In this Section, we shall discuss the consequences of the Bohr-van
Leeuwen theorem for the much debated problem of the thermal
Casimir effect in metallic systems. Therefore, we consider   the
Casimir apparatus consisting  of two non-magnetic plane-parallel
slabs, separated by a vacuum gap of width $d$.  We let
$\epsilon_1(\omega)$ and $\epsilon_2(\omega)$ the permittivities
of the two slabs,  which are assumed to occupy, respectively, the
regions $z \le 0$ and $z \ge d$.

The   Green's functions for this problem were worked out in Ref.
\cite{bimonte}, and we can use them here.  The material
contribution ${\tilde {\bf F}}^{(\rm cav)}_{\perp}$ to the
transverse Green's function was found to be of the form: \be
{\tilde {\bf F}_{\perp}}^{({\rm cav})}={\tilde {\bf U}}^{({\rm
cav})}+{\tilde {\bf V}}^{({\rm cav})}\;.\label{Gcavsplit}\ee Here
\begin{widetext}$$ {\tilde { U}_{ij}}^{({\rm cav})}= i \int \frac{d^2 {\bf k}_{\perp}}{{2
\pi}{k_z}} \left\{ \left[ \left(\frac{1}{{\cal A}_s}-1
\right)\left(e^{i { {\bf k}}^{(+)}\cdot ({\bf r}-{\bf r}')} + e^{i
{{\bf k}}^{(-)}\cdot ({\bf r}-{\bf r}')}\right) +
\frac{{r}_1^{(s)}}{{\cal A}_s}   \,e^{i { {\bf k}}^{(+)}\cdot {\bf
r} -i {{\bf k}}^{(-)}\cdot{\bf r}'}+ \frac{{r}_2^{(s)}}{{\cal
A}_s}\,e^{i { {\bf k}}^{(-)}\cdot {\bf r} -i { {\bf
k}}^{(+)}\cdot{\bf r}'+2 i k_z d}\right]   e_{\perp i} e_{\perp j}
\right.$$ $$ \left.+\frac{1}{k_0^2}\left[\left(\frac{1}{{\cal
A}_p}-1 \right)\left(\xi^{(+)}_i \xi^{(+)}_j e^{i { {\bf
k}}^{(+)}\cdot ({\bf r}-{\bf r}')} +\xi^{(-)}_i \xi^{(-)}_j e^{i {
{\bf k}}^{(-)}\cdot ({\bf r}-{\bf r}')}\right) + \xi^{(+)}_i
\xi^{(-)}_j\,\frac{r_{1}^{(p)}}{{\cal A}_p}  \,e^{i {\bf
k}^{(+)}\cdot {\bf r}-i{\bf k}^{(-)}{\bf r}' } \right.
\right.$$\be \left.\left. + \,\xi^{(-)}_i
\xi^{(+)}_j\,\frac{r_{2}^{(p)}}{{\cal A}_p} \,e^{i {\bf
k}^{(-)}\cdot {\bf r}-i{\bf k}^{(+)}{\bf r}' + 2 i k_z
d}\frac{}{}\right]\right\}\;,\label{Ucav} \ee
\end{widetext}
where $r_i^{(\alpha)},\;\alpha=s,p$ are the Fresnel reflection
coeffcients of slab $i$ for polarization $\alpha$, and ${\cal
A}_{\alpha}=1-r_1^{(\alpha)}r_2^{(\alpha)}\,\exp(2 i k_z d)$. As
to ${\tilde {\bf V}}^{({\rm cav})}$ it has the expression: \be
{\tilde {V}}_{ij}^{({\rm cav})}=\frac{1}{k_0^2} \frac{\partial^2 {
\Psi}^{({\rm cav})}}{\partial x_i \partial x'_j}\;,\label{Vcav}\ee
with
$$ {\tilde {\Psi}}^{({\rm cav})} = \int \frac{d^2 {\bf
k}_{\perp}}{2 \pi \,k_{\perp}}\left[ \left(\frac{1}{{\cal A}}-1
\right)\left(e^{i {\bar {\bf k}}^{(+)}\cdot ({\bf r}-{\bf r}')} +
e^{i {\bar {\bf k}}^{(-)}\cdot ({\bf r}-{\bf r}')}\right)\right.$$
\be \left. - \frac{1}{{\cal A}}\left( {\bar r}_1 \,e^{i {\bar {\bf
k}}^{(+)}\cdot {\bf r} -i {\bar {\bf k}}^{(-)}\cdot{\bf r}'}+{\bar
r}_2\,e^{i  {\bar {\bf k}}^{(-)}\cdot {\bf r} -i {\bar {\bf
k}}^{(+)}\cdot{\bf r}'-2 k_{\perp} d}
\right)\right]\;,\label{psicav}\ee where ${\cal A}=1-{\bar
r}_1(\omega)\,{\bar r}_2(\omega)\,\exp (-2 k_{\perp}\,d)$.
Concerning the matter contribution to the scalar Green's function
${\tilde { F}}^{(\rm cav)}$, in Ref. \cite{bimonte} it was shown
that \be {\tilde { F}}^{(\rm cav)}={\tilde {\Psi}}^{({\rm
cav})}\;.\ee As it is well known \cite{lifs} the Casimir pressure
acting on the slabs is given by the thermal average of the $zz$
component $\langle T_{zz}^{(\rm mat)}\rangle$ of the matter
contribution to the Maxwell stress-tensor. Since, according to Eq.
(\ref{UA}) the scalar potential and the vector potential are
uncorrelated, $\langle T_{zz}^{(\rm mat)}\rangle$ decomposes into
the sum of a longitudinal and a transverse contributions: \be
\langle T_{zz}^{(\rm mat)}\rangle=\langle T_{\| zz}^{(\rm
mat)}\rangle+\langle T_{\perp zz}^{(\rm mat)}\rangle\;,\ee and it
is interesting to evaluate the two contributions separately. Upon
recalling the classical expression of the Maxwell stress tensor in
vacuum: \be T_{ij}=\frac{1}{4 \pi}\left(\frac{1}{2}\delta_{ij} E_k
E_k-E_i\,E_j+\frac{1}{2}\delta_{ij} B_k B_k-B_i\,B_j\right)\;,\ee
we find \be\langle T_{\| zz}^{(\rm mat)}\rangle=\frac{1}{8
\pi}\sum_{k } \lambda_k \,\delta\langle\{E_{\| k}({\bf
r},0)\,E_{\| k}({\bf r},0)\}\rangle\;,\ee and $$\langle T_{\perp
zz}^{(\rm mat)}\rangle=\frac{1}{8 \pi}\sum_{k }
 \lambda_k \,\left[\delta\langle\{E_{\perp k}({\bf r},0)\,E_{\perp
k}({\bf r},0)\}\rangle \right.$$\be\left.+ \,\delta\langle\{B_{
k}({\bf r},0)\,B_{ k}({\bf r},0)\}\rangle\right]\;,\ee where
$\lambda_1=\lambda_2=-\lambda_3=1$. By recalling Eqs.
(\ref{ElEl}), (\ref{EE}) and (\ref{BB}) and using the explicit
expression of ${\tilde { F}}^{(\rm cav)}$ and  ${\tilde {
F}}_{\perp ij}^{(\rm cav)}$ given above, after some elementary
algebra one obtains the following expressions for $\langle T_{\|
zz}^{(\rm mat)}\rangle$ and $\langle T_{\perp zz}^{(\rm
mat)}\rangle$: $$ \langle T_{\| zz}^{(\rm
mat)}\rangle=\frac{1}{\pi^2}\int_0^{\infty}
\!\!\frac{d\omega}{\omega} E_{\beta}(\omega)\!\int {d{k}_{\perp}}
\,k_{\perp}^2$$ \be \times \, {\rm Im}\left[\left(1- \frac{e^{2
k_{\perp} d}}{{\bar r}_1 {\bar r}_2}\right)^{-1}\right],\ee \be
\langle T_{\perp zz}^{(\rm
mat)}\rangle=-\frac{2}{\pi}\int_0^{\infty}
\!\!\frac{d\omega}{\omega} E_{\beta}(\omega)\,{\rm Im}\, [{\cal
T}_{\perp}(\omega)]\,\label{Tzztra}\ee where
$$ {\cal T}_{\perp}(\omega)= \frac{1}{2 \pi}\!\int_0^{\infty}
{d{k}_{\perp}} \,k_{\perp} \left\{q \sum_{\alpha=s,p}\left(
\frac{e^{-2i k_{z} d}}{{r}_1^{(\alpha)}
{r}_2^{(\alpha)}}-1\right)^{-1}\right.$$\be\left.-
\,k_{\perp}\left( \frac{e^{2 k_{\perp} d}}{{\bar r}_1 {\bar
r}_2}-1\right)^{-1}\right\}\;, \label{calT}\ee with $q=-i k_z$. We
note that when we take the sum of $\langle T_{\| zz}^{(\rm
mat)}\rangle$ and $\langle T_{\perp zz}^{(\rm mat)}\rangle$, the
longitudinal contribution $\langle T_{\| zz}^{(\rm mat)}\rangle$
cancels against the transverse contribution resulting from the
second term between the curly brackets on the r.h.s. of Eq.
(\ref{calT}). The resulting expression for $\langle T_{zz}^{(\rm
mat)}\rangle$ then reproduces the following well known Lifshitz
formula for the Casimir pressure $P(d,T)$, expressed as an
integral over real frequencies \cite{lifs}: $$
P(d,T)=-\frac{1}{\pi^2}\int_0^{\infty} \!\!\frac{d\omega}{\omega}
E_{\beta}(\omega)\, \int_0^{\infty} {d{k}_{\perp}} \,k_{\perp}
$$
\be \times \,{\rm Im}\,\left\{q \sum_{\alpha=s,p}\left(
\frac{e^{-2i k_{z} d}}{{r}_1^{(\alpha)}
{r}_2^{(\alpha)}}-1\right)^{-1}\right\}\,.\label{lifs}\ee

We can now verify whether the transverse contribution $\langle
T_{\perp zz}^{(\rm mat)}\rangle$ vanishes in the classical limit,
as required by the Bohr-van Leeuwen theorem. By taking the limit
of the r.h.s. of Eq. (\ref{Tzztra}) for $\hbar \rightarrow 0$ we
obtain: \be \lim_{\hbar \rightarrow 0} \langle T_{\perp zz}^{(\rm
mat)}\rangle=-\frac{2 k_B T}{\pi}\,{\rm Im }\,\int_0^{\infty}
\!\!\frac{d\omega}{\omega}\, {\cal
T}_{\perp}(\omega)\;.\label{Tperpzzcl}\ee Evaluation of the
integral on the r.h.s. is made easy after we note that the
quantity ${\cal T}_{\perp}(\omega)$ vanishes for large
frequencies, and is an analytic function of the frequency in the
upper complex half-plane ${\cal C}^+$, as a result of analyticity
properties of the reflection coefficients. These properties permit
us to rotate the contour of integration from the real axis towards
the imaginary axis, as we did already in Section II. Since along
the imaginary axis ${\cal T}_{\perp}(w)$ is  real (because the
reflection coefficients are real for imaginary frequencies), one
finds that the integral on the r.h.s. of Eq. (\ref{Tperpzzcl})
receives its only contribution from the pole in the origin. The
latter is easily evaluated, giving us the   result: \be
\lim_{\hbar \rightarrow 0} \langle T_{\perp zz}^{(\rm
mat)}\rangle=- \,k_B T\, \lim_{\omega \rightarrow 0}\, {\cal
T}_{\perp}(\omega)\;.\label{Tperpzzclbis}\ee We then reach again
the conclusion that validity of the Bohr- van Leeuwen theorem
depends on the behavior of the reflection coefficients, or what is
the same of the permittivities, in the limit of zero frequency. At
this point we assume for simplicity that the slabs are made of the
same material, in such a way that
$\epsilon_1(\omega)=\epsilon_2(\omega)=\epsilon(\omega)$, and we
consider again the four models Eqs. (\ref{ins}-\ref{ideal}) for
the permittivity in the low-frequency limit. It is easy to check
that for all these models the quantity: \be q \left( \frac{e^{-2i
k_{z} d}}{({r}^{(p)})^2}-1\right)^{-1}- \,k_{\perp}\left(
\frac{e^{2 k_{\perp} d}}{{\bar r}^2}-1\right)^{-1}\ee occurring on
the r.h.s. of Eq. (\ref{calT}), vanishes like $\omega^2$, and
therefore it does not contribute to ${\cal T}_{\perp}$ in the
limit of vanishing frequencies. In view of this, only   TE-modes
may give a contribution to ${\cal T}_{\perp}$ in the limit of
vanishing frequencies and   from Eq. (\ref{calT}) we obtain:
$$ \lim_{\hbar \rightarrow 0} \langle T_{\perp zz}^{(\rm
mat)}\rangle=- \frac{\,k_B T}{2 \pi}\,\int_0^{\infty}
{d{k}_{\perp}} \,k^2_{\perp}$$ \be \times\, \left( \frac{e^{2
k_{\perp} d}}{({r}^{(s)}(0,k_{\perp}))^2}
 -1\right)^{-1}\;.\label{TEclas}\ee  The Bohr-van Leeuwen  theorem
 requires that the quantity on the r.h.s. vanishes for all separations, and
 this is only possible if ${r}^{(s)}(0,k_{\perp})$ vanishes. Recalling  the
 values for TE reflection
coefficient ${r}^{(s)}(0,k_{\perp})$   listed in Eqs.
(\ref{rTEins}-\ref{rTEid}), we see again that only the insulator
and  Drude-like models are consistent with the Bohr- van Leeuwen
theorem, while  plasma-like models and the ideal  model are not.

As it is well known after rotation of the frequency domain of
integration from the real axis to the imaginary axis, Lifshitz
formula Eq. (\ref{lifs}) takes the form of a sum over so-called
imaginary Matsubara frequencies: $$ P(d,T)=-\frac{k_B
T}{\pi}\sum_{n=0}^{\infty}\left(1-\frac{1}{2}\delta_{n0}\right)\int_0^{\infty}
{d{k}_{\perp}} \,k_{\perp}\,q_n
$$
\be \times  \, \sum_{\alpha=s,p}\left( \frac{e^{2 q_n
d}}{{r}_1^{(\alpha)}(i \xi_n,k_{\perp})\, {r}_2^{(\alpha)}(i
\xi_n,k_{\perp})}-1\right)^{-1} \,,\label{lifsim}\ee where
$\xi_n=2 \pi n k_B T/\hbar$ are the Matsubara frequencies and
$q_n=\sqrt{k_{\perp}^2+\xi_n^2/c^2}$. By comparing the r.h.s. of
Eq. (\ref{TEclas}) with the r.h.s. of Lifshitz formula Eq.
(\ref{lifsim}) we see that the former quantity coincides with the
$n=0$ contribution to the Casimir pressure for TE polarization. As
it is well known this very term has been the object of a long
debate in the recent literature on the thermal Casimir effect, and
as of now there is no consensus among experts on its actual
magnitude in the case of normal metallic plates
\cite{Mohid,brevik,qfext}. The analysis carried here shows that
the Bohr-van Leeuwen theorem requires that this term be zero for
normal non magnetic conductors. Quantum effects permit of course
non vanishing values for the $n=0$ TE contribution, without
violating the theorem, for example in magnetic materials and
especially in superconductors \cite{bimonte2} (see remarks at the
end of the previous Section).

\section{CONCLUSIONS AND DISCUSSION}

The problem of the thermal correction to the Casimir and
Casimir-Polder interactions in systems involving normal ohmic
conductors has attracted considerable attention in the recent
literature on dispersion forces. Despite numerous theoretical and
experimental investigations, the resolution of this problem is not
clear yet.

On the theoretical side, several distinct models have been
proposed, that give widely different predictions for the magnitude
of this correction. The difficulty is that in order to get a
definite value for the thermal correction, it is necessary to make
a definite extrapolation to zero frequency of the optical data of
the plates, because   the theoretical prediction is very sensitive
to the limiting behavior of the reflection coefficients for
vanishing frequency. Unfortunately, this extrapolation cannot be
done solely on the basis of direct measurements, and it
unavoidably requires making some theoretical assumptions.
Different ansatz have been proposed in the literature, each
supported by some arguments, that give widely different
predictions for the magnitude of the thermal correction.

The experimental situation is not definite either. In principle,
the best way to clarify the issue would be to measure the Casimir
force between two metallic plates at separations of a few microns,
because for such large separations different theoretical models
predict forces that differ by as much as about fifty percent.
Unfortunately, at these distances the Casimir force is very small,
and at the moment no one has been able to measure it with
sufficient precision \cite{brown}. Below one micron, where the
Casimir force can be measured most accurately, the thermal
correction gets very small and therefore it is difficult to
observe it. At the moment of this writing, the most relevant
experiments in this regard are those at Purdue University
\cite{decca} (see also Ref.\cite{Mohid}). While the achieved
precision is not sufficient to detect the thermal correction
predicted by certain theoretical models (of the plasma type), the
authors claimed that their measurements  rule out at high
confidence level the much larger thermal correction predicted by
alternative theoretical models (of ohmic conductor type). As of
now, no other experimental groups have carried out measurements of
the thermal correction to the Casimir force. For a definitive
assessment, it would be desirable to have more experiments,
possibly of different types. Perhaps, further insights may come
from superconductors \cite{bimonte2}.

In this paper we have examined the most widely used models for the
low-frequency behavior of the reflection coefficients of
dielectrics and conductors, from the point of view of their
consistency with the Bohr- van Leeuwen theorem \cite{van}. This is
a well known theorem  in classical statistical physics, stating
that at thermal equilibrium the transverse electromagnetic field
decouples from matter. As it is well known, this theorem provides
the basic explanation for the absence of strong diamagnetism in
normal conductors. The theorem has a very general character, and
it fact it also holds under the less restrictive assumption of
local kinetic equilibrium \cite{roth}. In this paper, we evaluated
the correlation functions for the transverse electromagnetic field
outside a dielectric or conducting slab, and inside a planar
cavity of the type used in Casimir experiments. Upon taking the
classical limit of these correlators, we found that decoupling of
the transverse electromagnetic field occurs if and only if the
reflection coefficient for transverse electric (TE) modes vanishes
at zero frequency, in the classical limit. According to this
result, we conclude that the dielectric and the ohmic conductor
models are completely consistent with the Bohr- van Leeuwen
theorem, while neither   plasma-like models nor the ideal metal
model are. Interestingly, in the case of a cavity, the average
value of the Maxwell stress tensor, providing the Casimir
pressure, is consistent with the theorem if and only if the $n=0$
Matsubara mode for TE polarization gives no contribution to the
Casimir pressure. This term is precisely the source of the present
controversies on the thermal Casimir effect for metallic plates.
From the point of view of the Bohr- van Leeuwen theorem the answer
is clear: this term must be zero, up to quantum effects. We remark
that non-vanishing values for the TE reflection coefficient are of
course possible in materials displaying a magnetic response, and
in superconductors, since  both phenomena arise from quantum
effects that  disappear in the classical limit. When such
materials are considered a on-vanishing contribution to the
Casimir pressure from the $n=0$ Matsubara mode is clearly
possible, without violating the Bohr- van Leeuwen theorem.

The reader may be disturbed by hearing that the familiar
ideal-metal model is inconsistent with the Bohr- van Leeuwen
theorem. After all, this is the model that Casimir himself used to
investigate the effect that goes under his name, sixty years ago
\cite{casimir}. We remark that this model constitutes an extreme
idealization of real metals, and therefore one should not expect
it to be universally valid. A hint that this model may provide
incorrect answers in problems involving proximity effects of the
quantized electromagnetic field near material surfaces was pointed
out long ago by Milonni \cite{milonni} who found that ideal-metal
boundary conditions imply a violation of equal-time canonical
commutation relations for the transverse electromagnetic field in
the vicinity of a metallic surface. In a recent work
\cite{bimonte}, the author of the present paper showed that such a
violation is an artifact of the idealized boundary conditions for
a perfect metal. Indeed, if account is taken of dispersive and
dissipative features of real materials, no such violation is
found, and the canonical commutation relations are restored at all
point outside the conductor. As to  the range of validity of the
ideal-metal model, we have seen above that the field correlators
derived from this model violate the Bohr- van Leeuwen theorem only
at finite temperature and only in the vicinity of the metallic
surface. We infer from this that the model has to be considered
unreliable in the study of the proximity effects like the Casimir
and the Casimir-Polder interactions, in situations where
temperature effects become important.

Finally, we note that in the recent literature on the Casimir
effect there has been a lively debate on the influence of free
charge carriers that always exist in poor conductors and in
semiconductors \cite{Mohid,lamor}. The Bohr- van Leeuwen theorem
cannot give any help   to discriminate between the alternative
models that have been proposed for these materials, because in
this case differences in the predicted magnitudes of the Casimir
and Casimir-Polder interactions arise either from longitudinal
fields or from TM fields. As far as longitudinal fields are
concerned, the Bohr- van Leeuwen theorem obviously does not apply.
In the case of TM electromagnetic fields, we have seen in Sections
IV and V that they always decouple in the classical limit,
whatever model is used for the dielectric function.


\begin{thebibliography}{200}




\bibitem{parse} V. A. Parsegian, {\it Van der Waals Forces} (Cambridge University Press, Cambridge, England, 2005).

\bibitem{lifs} E.M. Lifshitz, Sov. Phys. JETP {\bf 2}, 73 (1956);
E.M. Lifshitz and L.P. Pitaevskii, {\it Landau and Lifshitz Course
of Theoretical Physics: Statistical Physics Part II}
(Butterworth-Heinemann, 1980).

\bibitem{Mohid} G.L. Klimchitskaya, U. Mohideen and V.M.
Mostepanenko,
arXiv:0902.4022.

\bibitem{cornell} J.M. Obrecht, R.J. Wild, M. Antezza, L.P.
Pitaevskii, S. Stringari and E.A. Cornell, Phys. Rev. Lett. {\bf
98}, 063201 (2007).

\bibitem{iannuzzi} W.J. Kim, M. Brown-Hayes, D.A.R. Dalvit, J.H. Brownell and R. Onofrio,
Phys. Rev. A {\bf 78}, 020101(R) (2008); A {\bf 79}, 026102
(2009); R.S. Decca et al., ibid. A {\bf 79}, 026101 (2009); S. de
Man, K. Heeck and D. Iannuzzi, ibid. A {\bf 79}, 024102 (2009).

\bibitem{piro}  I. Pirozhenko, A.
Lambrecht and V.B. Svetovoy, New J. Phys. {\bf 8}, 238 (2006).

\bibitem{sernelius}   M. Bostrom and B.E. Sernelius,
Phys. Rev. Lett. {\bf 84}, 4757 (2000); B.E. Sernelius, {\it
ibid.} {\bf 87}, 139102 (2001).

\bibitem{geyer} B. Geyer, G.L. Klimchitskaya and V.M.
Mostepanenko, J. Phys. A {\bf 40}, 13485 (2007).

\bibitem{brevik} I. Brevik,  S.A. Ellingsen and
K.A. Milton, New J. Phys. {\bf 8}, 236 (2006).

\bibitem{qfext} {\it Proccedings of the 8th Workshop on Quantum Field Theory
under the Influence of External Conditions} ({\rm edited by M.
Bordag and V.M. Mostepanenko}) J. Phys. A {\bf 41},
 (2008).

\bibitem{brown} M. Brown-Hayes {\it et al.}, J. Phys. A {\bf 39}, 6195
(2006); P. Antonini {\it et al.}, arXix:0812.5065.

\bibitem{decca} R.S. Decca, D. Lopez, E. Fischbach, G.L. Klimchitskaya, D.E. Krause and V.M.
Mostepanenko, Ann. Phys. {\bf 318}, 37 (2005); Eur. Phys. J. C
{\bf 51}, 963 (2007), Phys. Rev. D {\bf 75}, 077101 (2007).

\bibitem{Iannuzzi} S. de Man, K. Heeck, R. J. Wijngaarden and D.
Iannuzzi, arXiv:0901.3720.

\bibitem{van} H. J., van Leeuwen, J. Phys. Radium {\bf 2}, 362 (1921);
J. H. Van Vleck, {\it The Theory of Electric and Magnetic
Susceptibilities} (Clarendon Press, Oxford, 1932).

\bibitem{callen} H. B. Callen and T. A. Welton, Phys. Rev {\bf
83}, 34 (1951); R. Kubo, Rep. Prog. Phys. {\bf 29}, 255 (1966); R.
Kubo, in {\it Statistical Mechanics of Equilibrium and
Non-Equilibrium Systems}, edited by J. Meixner (North-Holland,
Amsterdam,1965), p. 81.

\bibitem{Casimir2} H.B.G. Casimir, Rev. Mod. Phys. {\bf 17}, 343
(1945).

\bibitem{santa} G. Bimonte and E. Santamato, Phys. Rev. A {\bf 76},
013810 (2007).

\bibitem{agarwal} G. S. Agarwal, Phys. Rev. A {\bf 11}, 230, 243, 253 (1975); A {\bf 12}, 1475,
1974 (1975).

\bibitem{Martin} P. R. Buenzli and Ph. A. Martin, Phys. Rev. E {\bf
77}, 011114 (2008).

\bibitem{bezerra} V.B. Bezerra, G.L. Klimchitskaya and V.M.
Mostepanenko, Phys. Rev. A {\bf 65}, 052113 (2002); {\bf 66},
062112 (2002);  V.B. Bezerra, G.L. Klimchitskaya, V.M.
Mostepanenko and C. Romero, {\it ibid.} {\bf 69} 022119 (2004).

\bibitem{sernelius2} M. Bostr{\"{o}m} and B.E. Sernelius, Physica
A {\bf 339}, 53 (2004).

\bibitem{svetovoy} V.B. Svetovoy and M.V. Lokhanin, Phys. Rev. A {\bf 67}, 022113 (2003),
V.B. Svetovoy and R. Esquivel, Phys. Rev. E {\bf 72}, 036113
(2005).

\bibitem{elling} S.A. Ellingsen, I. Brevik, J.S. H{\o}ye and
K.A. Milton, quant-ph:0809.0763.

\bibitem{bimonte2} G.Bimonte, E. Calloni, G. Esposito, L. Milano and L.
Rosa, Phys. Rev. Lett. {\bf 94}, 180402 (2005); G. Bimonte, E.
Calloni, G. Esposito, and L. Rosa, Nucl. Phys. B {\bf 726}, 441
(2005); G. Bimonte {\it et al.},   J. Phys. A {\bf 41}, 164023
(2008); G. Bimonte, Phys. Rev. A {\bf 78}, 062101 (2008).

\bibitem{roth} J. R. Roth, Nasa Tech. Note, Nasa TN D-3880 (1967).

\bibitem{casimir}  H.B.G.~Casimir, Proc. K. Ned. Akad. Wet.
Rev. {\bf 51}, 793 (1948).

\bibitem{milonni} P. W. Milonni, Phys. Rev. A {\bf 25}, 1315
(1982).

\bibitem{bimonte} G. Bimonte, arXiv:0902.4537.

\bibitem{lamor} L.P. Pitaevskii, Phys. Rev. Lett.
{\bf 101}, 163202 (2008); D.A.R. Dalvit and S.K. Lamoreaux, ibid.
{\bf 101}, 163203 (2008); V.B. Svetovoy, ibid. {\bf 101}, 163603
(2008).











\end{thebibliography}
\end{document}